\documentclass[fleqn,usenatbib]{mnras}

\usepackage{newtxtext,newtxmath}
\usepackage{setspace}

\usepackage[T1]{fontenc}

\DeclareRobustCommand{\VAN}[3]{#2}
\let\VANthebibliography\thebibliography
\def\thebibliography{\DeclareRobustCommand{\VAN}[3]{##3}\VANthebibliography}

\usepackage{graphicx}	
\usepackage{amsmath}	
\usepackage{float}      
\usepackage{booktabs}   
\usepackage{comment}
\usepackage{anyfontsize}

\title[]{Color Dependence of  Dipole in CatWISE2020 Data}

\author[P. Mohit et al.]{
Mohit Panwar,$^{1}$\thanks{E-mail: mohitpan@iitk.ac.in}
Pankaj Jain,$^{2}$\thanks{E-mail: pkjain@iitk.ac.in}
Amitesh Omar$^{2}$
\\
$^{1}$Department of Physics, Indian Institute of Technology, Kanpur 208016, India\\
$^{2}$Department of Space, Planetary \& Astronomical Sciences \& Engineering (SPASE), Indian Institute of Technology, Kanpur 208016, India
}

\date{Accepted XXX. Received YYY; in original form ZZZ}

\pubyear{2024}

\begin{document}
\label{firstpage}
\pagerange{\pageref{firstpage}--\pageref{lastpage}}

\maketitle

\begin{abstract}
The signal of dipole anisotropy in quasar number counts is studied using the CatWISE2020 catalog in various color bins. It is found that the dipole signal differs significantly in two color bins, namely, $1.1>W1-W2\ge 0.8$ and $1.4>W1-W2>1.1$. The color bin $1.4>W1-W2>1.1$ appears strongly contaminated, with possibly Galactic contributions and is unreliable for extracting the signal of cosmological dipole. The source of this contamination has not been identified and cannot be attributed to known emissions within the galaxy. Removing this contaminated color bin leads to a strong dipole signal with a direction significantly different from that obtained from full data. If we interpret this dipole as due to our local motion, the extracted velocity turns out to be $900\pm 113$ Kms$^{-1}$, which deviates from the CMB dipole velocity with approximately $4.7$ sigma significance.
\end{abstract}

\begin{keywords}
large-scale structure of Universe -- quasars:general -- cosmology:miscellaneous
\end{keywords}



\section{Introduction}
The standard $\Lambda$CDM model is based on the cosmological principle (CP), which asserts that the Universe is isotropic and homogeneous on a sufficiently large distance scale. The precise distance scale at which this applies is so far unknown, although it is generally believed to be of order $100$ Mpc. The CP naturally sets a preferred frame, also known as the cosmic rest frame (CRF), in which the cosmological observations appear isotropic and homogeneous. On theoretical grounds, the cosmological principle may be justified by inflation. The basic idea is that, at the beginning of time, the Universe may have a complicated structure, but during inflation, the observable part of the Universe becomes isotropic and homogeneous. This has been explicitly shown for a limited class of models, in which an initially anisotropic Universe acquires isotropy within the first e-fold of inflation \cite{PhysRevD.28.2118}. However, this hypothesis directly suggests a limitation of the cosmological principle. For example, the perturbative modes originating during or before the first e-fold need not obey this principle \cite{Aluri:2011xm,Rath:2013bfa}.

Observationally, there are some claims on the violation of the cosmological principle. The radio polarizations of distant galaxies show a dipole anisotropy with a preferred axis pointing close to the cosmic microwave background (CMB) dipole \cite{Jain1998,Ralston2004}. Subsequently, several effects have been found, showing an anisotropy correlated with this direction. This includes the quadrupole and octopole alignment in CMB \cite{Quadrupole_and_octopole_alignment}, hemispherical power asymmetry \cite{Eriksen_2004_hemispherical_power_asymmetry}, dipole in galaxy number counts \cite{1998MNRAS.297..545B, Blake_2002, Singal_2011, 10.1111/j.1365-2966.2012.22032.x, Rubart, TIWARI20151, Bengaly_2018, Secrest_2021}, large scale cosmic flow \cite{Kashlinsky_2008, 10.1093/mnras/stad1984} and anisotropy in Hubble constant \cite{PhysRevD.105.103510} etc.
It has been pointed out that a dipole anisotropy in galaxy number counts is expected in the $\Lambda$CDM model due to kinematic effects \cite{Baldwin_Ellis_1984} and is given by,
\begin{equation}
    \vec{D} = [2+x(1+\alpha)]\vec{\mathrm{v}}/c\,
\end{equation}
where $x$ is the index of cumulative number count $\mathcal{N}(>S)$, i.e., number of sources per unit solid angle having flux density greater than $S$, assumed to follow a power law $\mathcal{N}(>S)\propto S^{-x}$, $\alpha$ is the spectral index of the source's spectrum, defined by $S(\nu)\propto \nu^{-\alpha}$, $\vec{\mathrm{v}}$ is the peculiar velocity of the solar system with respect to the CRF, and $c$ is the speed of light in vacuum. The observed dipole in number counts, however, appears to be much larger than the kinematic expectation \cite{Singal_2011, 10.1111/j.1365-2966.2012.22032.x, Rubart, TIWARI20151, Bengaly_2018, Secrest_2021}.

An unusually strong departure from the $\Lambda$CDM prediction in the number count dipole, employing the CatWISE2020 \cite{Marocco_2021} catalog at infrared wavelengths, is claimed by \cite{Secrest_2021} with a significance of approximately $4.9$ sigma, and has also been confirmed by other authors \cite{2023MNRAS.525..231D, Kothari_2024}. It has been pointed out that the effect of the evolution of the luminosity function needs to be taken into account to assess properly the significance of the excess dipole amplitude \cite{Guandalin_2023}. Furthermore, no deviation from the $\Lambda$CDM model is seen in the Quaia data sample \cite{2023arXiv230617749S} by using Bayesian statistics \cite{2023MNRAS.tmp.3599M}. However, a recent analysis of the same Quaia data sample by \cite{Singal2024_Quaia_Dipole} reveals a significant deviation from the $\Lambda$CDM model, as the dipole amplitude is approximately $4$ times larger than that of the CMB-inferred dipole amplitude, and the direction is nearly similar to the CMB dipole direction.

In the present paper, the effect of color cut on the dipole signal in the quasar sample using the CatWISE2020 catalog \cite{Marocco_2021} is studied. \cite{Secrest_2021} imposed the color cut $1.4>W1-W2 \ge 0.8$ on W1 and W2 band magnitude data of the CatWISE2020 catalog. In order to study the dependence of the dipole signal on color $(W1-W2)$, the catalogue is divided into smaller bins of $(W1-W2)$, and dipole are extracted in each bin. If the signal is of cosmological origin, it should be independent of the color. 

The paper is organized in the following manner. In section \ref{sec:CatWISE2020_DATA}, we  describe the data used in this paper, the procedure used to eliminate the known systematics, and the corrections applied to remove the Galactic contamination. In section \ref{sec:methodology}, we discuss the method employed to estimate the dipole signal. The main results of this paper are given in section \ref{sec:results}, and we conclude in section \ref{sec:conclusion}. Throughout this paper, low color cut or $W1-W2<1.1$ stands for $1.1>W1-W2\ge 0.8$ and high color cut or $W1-W2>1.1$ stands for $1.4>W1-W2>1.1$ respectively. 

\section{CatWISE2020 Catalog}\label{sec:CatWISE2020_DATA}
\begin{figure} 
    \includegraphics[width=\columnwidth]{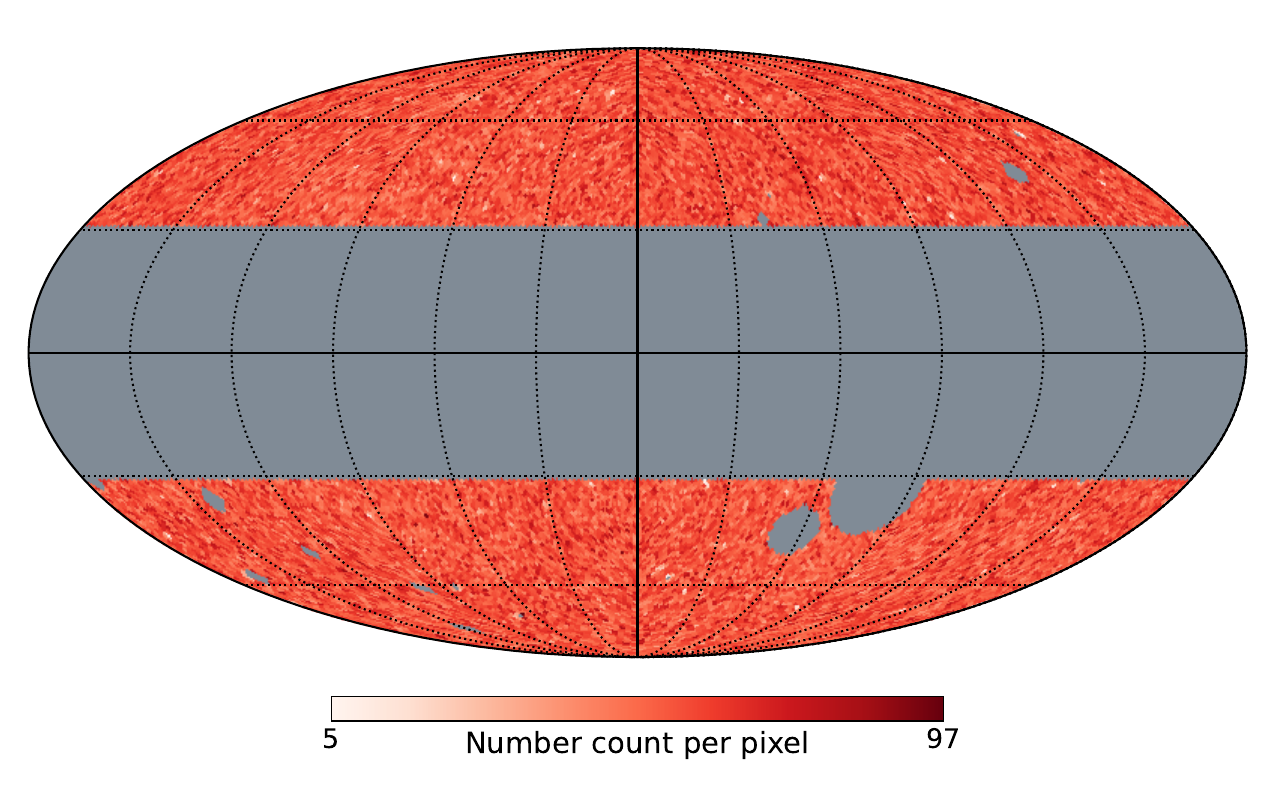}
    \caption{\label{fig:full_data_NumberCoutMap} Number count map using Mollweide projection in Galactic coordinate system. The masked regions are displayed in gray color.}
\end{figure}

We use a sample of $1355352$ quasars from the CatWISE2020 catalog \cite{Marocco_2021}, which is constructed using $\textbf{W}$ide-field $\textbf{I}$nfrared $\textbf{S}$urvey $\textbf{E}$xplorer (WISE) \cite{Wright_2010} and NEOWISE ($\textbf{N}$ear $\textbf{E}$arth $\textbf{O}$bject WISE) \cite{Mainzer_2011, Mainzer_2014} all-sky survey data at 3.4$\mu$m ($W1$ Band) and 4.6$\mu$m ($W2$ Band) wavelengths. A mid-infrared color cut $(W1-W2) \ge 0.8$ is applied to select quasars in the CatWISE2020 catalog \cite{Stern_2012, Secrest_2015}. Our quasar sample is identical to the sample used by \cite{Secrest_2021}, incorporating all cuts, masks, and corrections except for one specific correction, i.e., removing the observed inverse linear dependency of number density on absolute ecliptic latitude, believed to be caused by the WISE scanning pattern in the sky \cite{Wright_2010}. This correction is not essential since our analysis method can directly extract such dependency in the form of a quadrupole signal aligned with the ecliptic poles. The resulting quasar sample is the same as that used by \cite{Kothari_2024} and \cite{panwar2023probing}. To remove the Galactic contamination, the Galactic plane region $|b|<b_{cut}=30.0$ deg is masked. The magnitudes $W1$ and $W2$ are corrected for the Galactic reddening by using Planck dust map \cite{Planck_all_sky_thermal_dust_emission_model} and extinction coefficients $(A_{W1}/A_{V}=0.039\pm 0.004,\:\: A_{W2}/A_{V}=0.026\pm 0.004)$ derived in \cite{Wang_2019}. The ratio of total-to-selective extinction $R_{V}=A_{V}/E(B-V)$ is $3.1\pm 0.1$, where $A_{V}$ is extinction in standard Johnson visual band $(V)$ and $E(B-V)$ is reddening value in the line of sight of the source from the Planck dust map. We pixelize the quasars using a Python-based healpy package into equal-area pixels corresponding to nside$=64$. We imposed the constraint that the minimum number of sources in any pixel must be greater than or equal to $5$. A pixel which has smaller number of sources is masked. This constraint is applied throughout our analysis. The number count map is shown in Fig. \ref{fig:full_data_NumberCoutMap} in the Galactic coordinate system. All the masked regions are shown in gray color. We estimate the index of the integral number count $x$, using the log-likelihood method \cite{Ghosh_2017, panwar2023probing}, and the spectral index $\alpha$ for each source following \cite{Secrest_2021}.

\begin{figure}
    \includegraphics[width=\columnwidth]{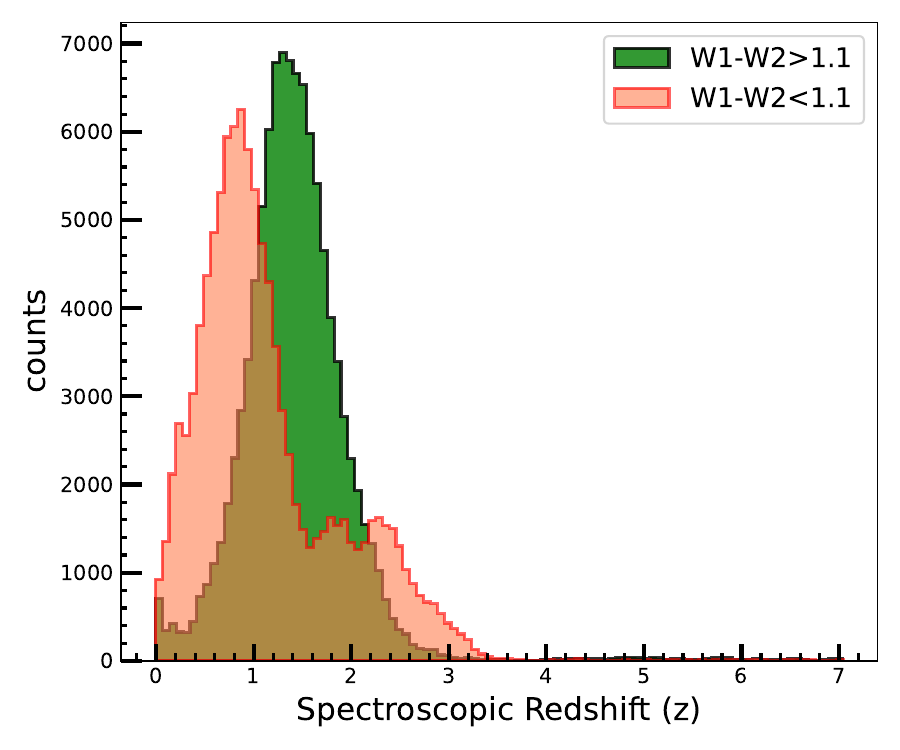}
    \caption{\label{fig:lowhighcolorcutredshiftDist.}The Spectroscopic Redshift$(z)$ distribution of low and high color cut sources. The average value of redshift for low and high color cut sources are $1.169$ and $1.435$ respectively. The spectroscopic classes of these sources are Quasar, Galaxy and Star. The fraction of these classes in the matched population are approximately $96\%$, $3\%$ and $0.2\%$ respectively. Hence our matched source population is mainly dominated by spectroscopically confirmed quasars.}
\end{figure}

We cross match our quasar sample with $\textbf{S}$loan $\textbf{D}$igital $\textbf{S}$ky $\textbf{S}$urvey (SDSS) data release $16$ \cite{SDSS_DR16_Ahumada_2020} quasar catalog \texttt{specObj-dr16.fits}\footnote{\url{https://data.sdss.org/sas/dr16/sdss/spectro/redux/specObj-dr16.fits}} within the search radius $3$ arcsec to extract the redshift$(z)$. 
Resulting distributions of $z$ for $W1-W2<1.1$ and $W1-W2>1.1$ sources are displayed in Fig. \ref{fig:lowhighcolorcutredshiftDist.}. 

\section{Methodology}\label{sec:methodology}
We obtain the dipole in number counts $N(\theta, \phi)$ by using the same procedure as in \cite{Kothari_2024}. We assume the following model for the angular dependence of number counts, 
\begin{eqnarray}
    N_{\mathrm{model}}(\theta, \phi)&=&M_{o}+n_{x}D_{x}+n_{y}D_{y}+n_{z}D_{z} \nonumber\\
    &+& Q_{xy}n_{x}n_{y}+Q_{zy}n_{y}n_{z}+Q_{xz}n_{x}n_{z}\nonumber\\
    &+& Q_{x^{2}-y^{2}}(n_{x}^{2}-n_{y}^{2}) + Q_{z^2}(3n_{z}^2-1)
    \label{eq:numbercount_sky_model}
\end{eqnarray}
where we keep terms up to $l= 2$, namely the monopole ($M_o$), dipole ($D_{x},D_{x},D_{z}$) and the quadrupole  $(Q_{xy},Q_{zy},Q_{xz},Q_{x^{2}-y^{2}} Q_{z^2})$. Here, $(n_{x}, n_{y}, n_{z})=(\sin\theta\cos\phi,\: \sin\theta\sin\phi,\: \cos\theta)$ are the components of the unit vector in the spherical coordinate system. Hence, the model contains a total of nine parameters which are extracted by using $\chi^2$ minimization.
The normalized dipole amplitude $D$ is defined as,
\begin{equation}
{D}=  \frac{\sqrt{{D}^2_x +{D}^2_y  + {D}^2_z}}{{M}_o}.
\end{equation}
We pixelize the sky using the \texttt{Healpy} pixelization. The number counts in pixel $i$ are denoted by $N_i$. The corresponding $\chi^2$ function is given by, 
\begin{equation}
    \chi^{2} = \sum_{i=1}^{N_{\mathrm{pix}}}\frac{[N_{i}-N_{i,m}]^{2}}{N_{i}}
\end{equation}
where $N_{\mathrm{pix}}$ is the total number of pixels. Assuming Poisson statistics, we set the error to be $\sigma_i = \sqrt{N_i}$. We also determine the change in results as the galactic mask is made more stringent.

\section{results and discussion}\label{sec:results}
\begin{figure}
    \includegraphics[width=\columnwidth]{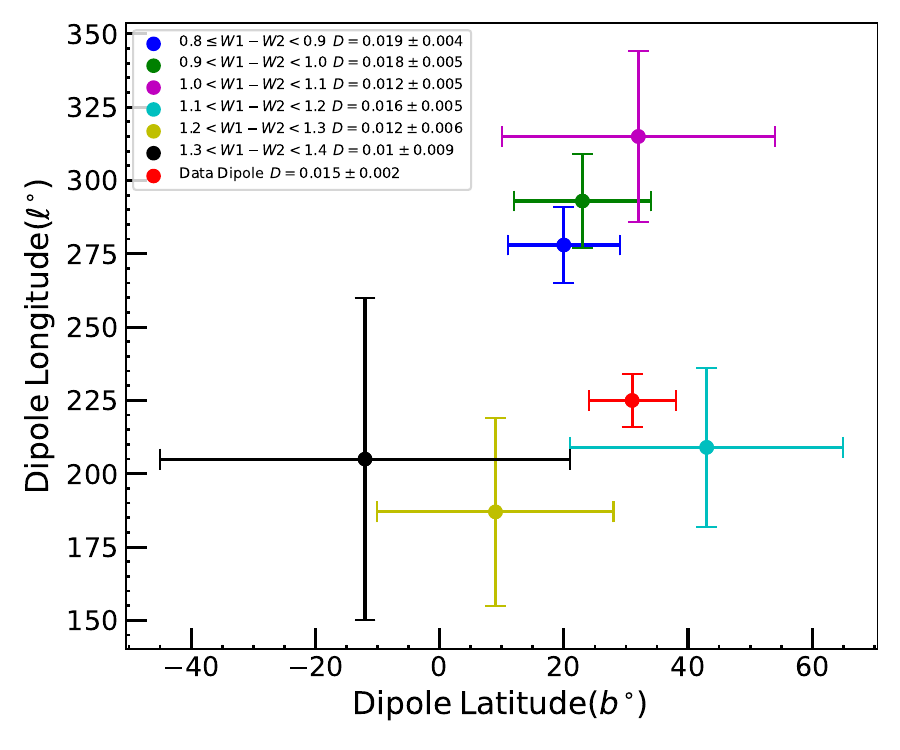}
    \caption{\label{fig:num_color_bin}A plot of the dipole longitude $(l^{\circ})$ vs dipole latitude $(b^{\circ})$ for different color bins. Dipole for full data sample is also shown in red color for comparison.}
\end{figure}

The extracted dipole parameters for the full data set are provided in Table \ref{table:fulldataDipoleResults}. The corresponding dipole parameters for various color cuts are given in Table \ref{table:2}. The dipole directions are plotted in Fig. \ref{fig:num_color_bin}. There is some difference in results between the two different sets of color cuts, with the dominant change being in the dipole directions. As shown in Fig. \ref{fig:num_color_bin}, the dipoles for the lower values of $W1-W2$ ($W1-W2<1.1$) cluster around the Galactic longitude of approximately $300^{\circ}$, while those for higher values of $W1-W2$ ($W1-W2>1.1$) are closer to $200^{\circ}$. We find that although the galactic latitudes are roughly in agreement, the galactic longitudes are very different. In Table \ref{table:highlowDipole}, we show the results after combining the lower and higher color cuts. The dipole for the high color cut $W1-W2>1.1$ points roughly opposite to the center of the galaxy, while that for $W1-W2<1.1$ is closer to the CMB dipole. We also find that the amplitude for $W1-W2>1.1$ is larger than that for $W1-W2<1.1$. 

\begin{figure} 
    \includegraphics[width=\columnwidth]{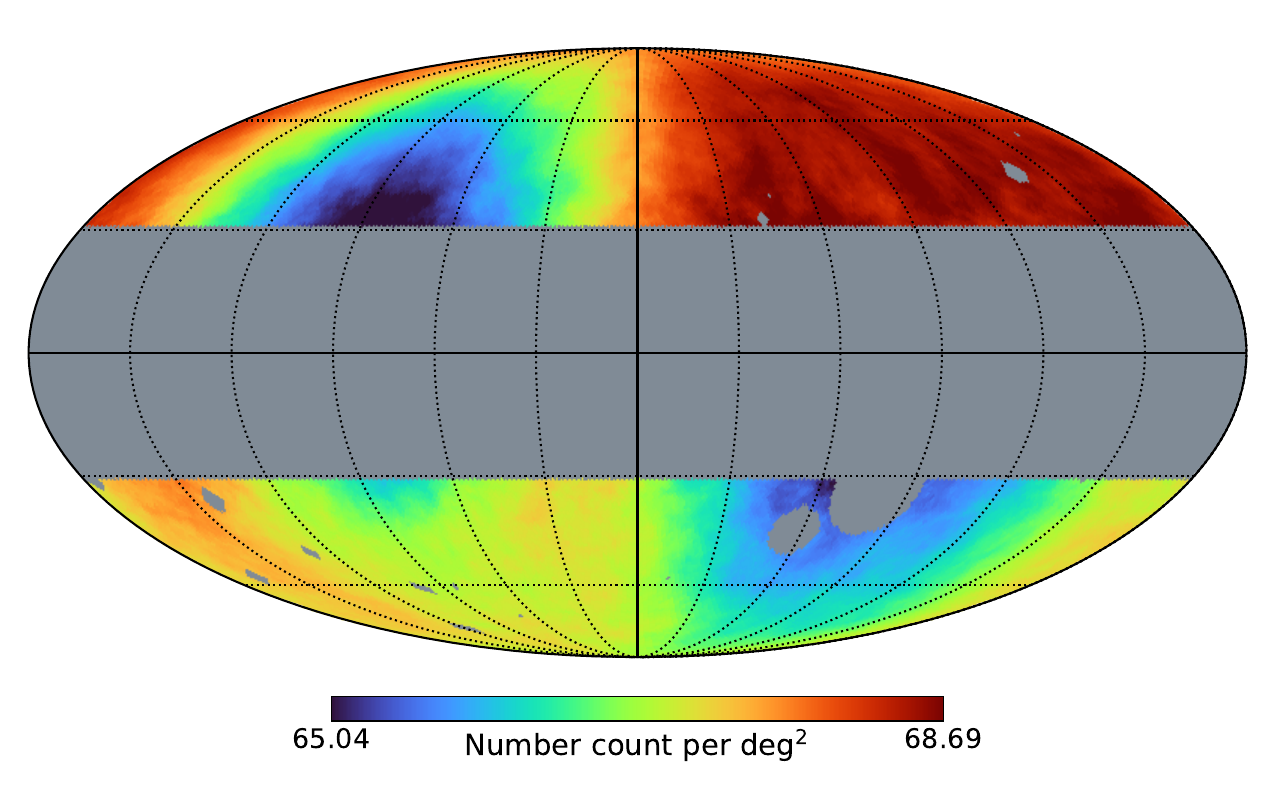}
    \caption{\label{fig:full_data_sky_distribution} Number density map i.e. number count per deg$^{2}$ obtained by smoothing the number count map Fig. \ref{fig:full_data_NumberCoutMap} using moving average on steradian scale. The masked regions are displayed in gray color.}
\end{figure}

\begin{figure}
    \includegraphics[width=\columnwidth]{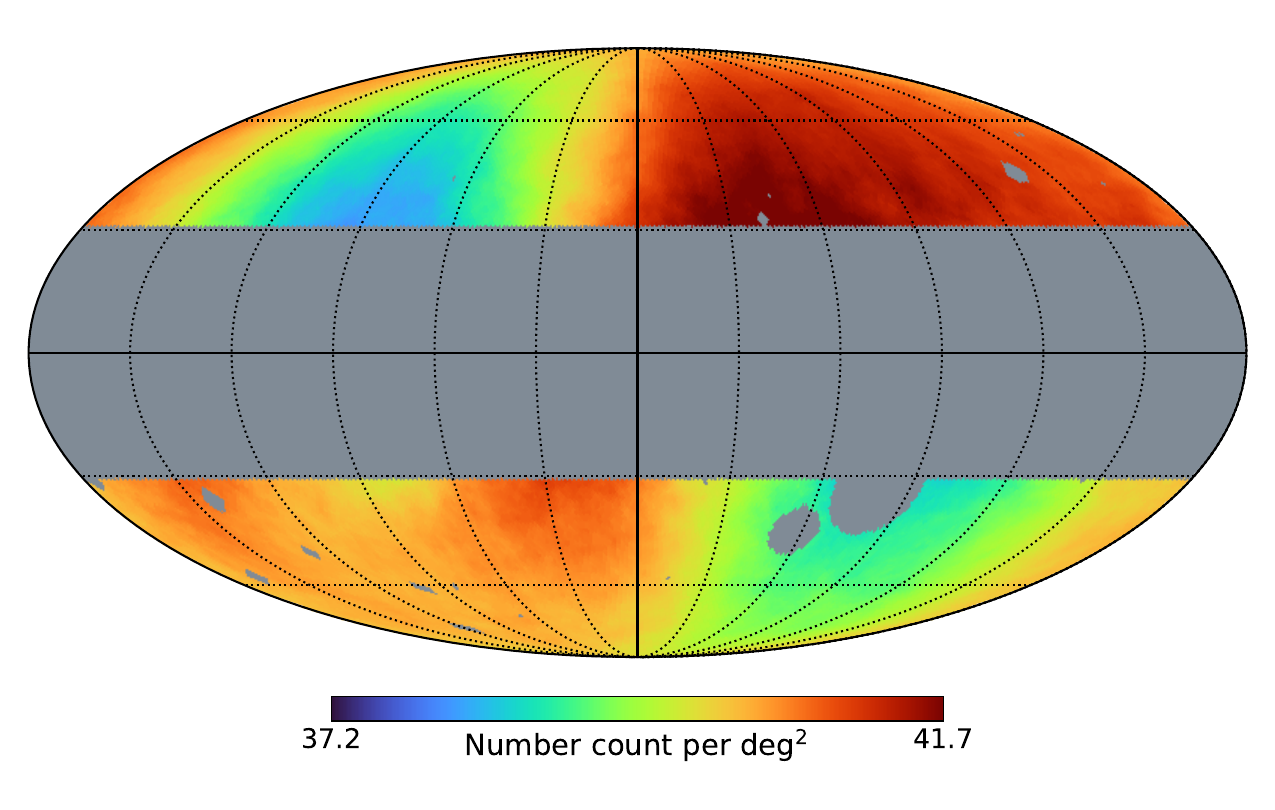}
    \caption{\label{fig:full_data_sky_distribution_low_color_cut} Number density map i.e. number count per deg$^{2}$ obtained by smoothing low color cut $(1.1>W1-W2\ge 0.8)$ data using moving average on steradian scale. The masked regions are displayed in gray color.}
\end{figure}

\begin{figure}
    \includegraphics[width=\columnwidth]{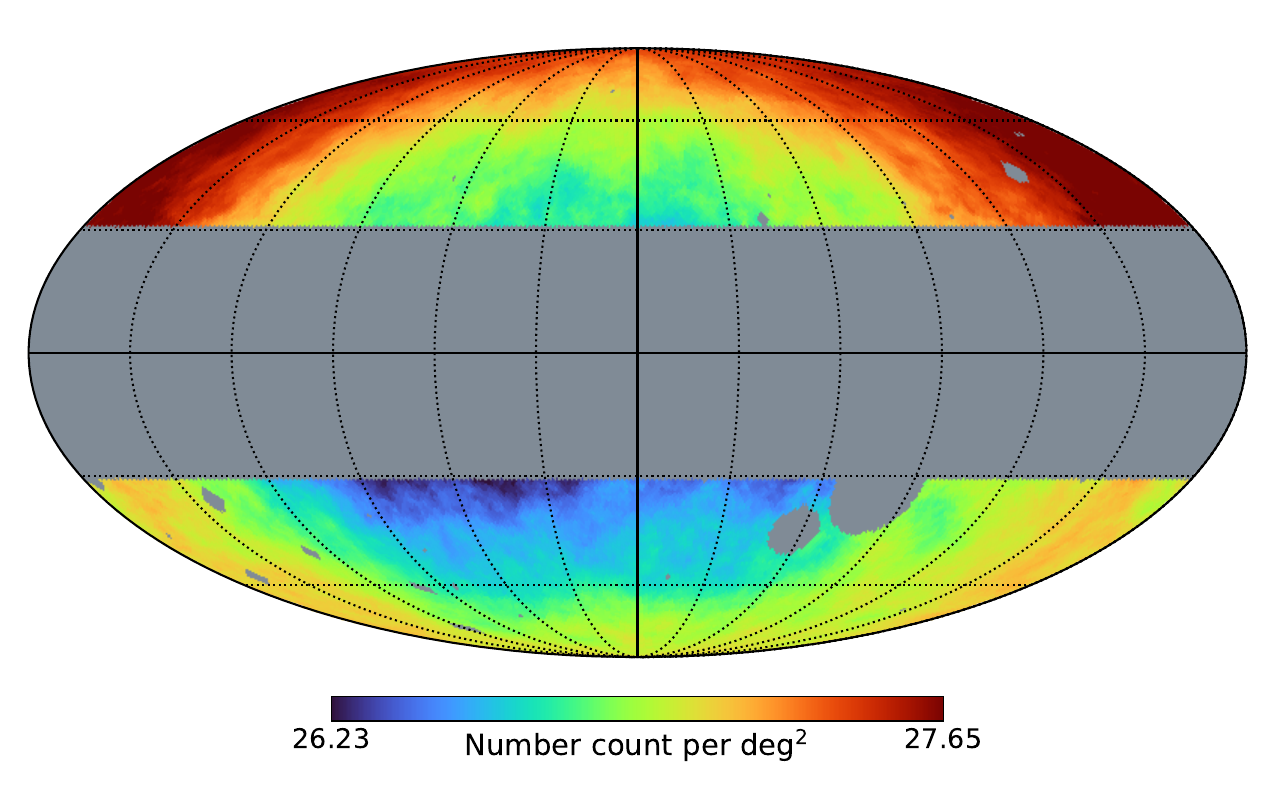}
    \caption{\label{fig:full_data_sky_distribution_high_color_cut} Number density map i.e. number count per deg$^{2}$ obtained by smoothing high color cut $(1.4>W1-W2>1.1)$ data using moving average on steradian scale. The masked regions are displayed in gray color.}
\end{figure}

\begin{figure}
    \includegraphics[width=\columnwidth]{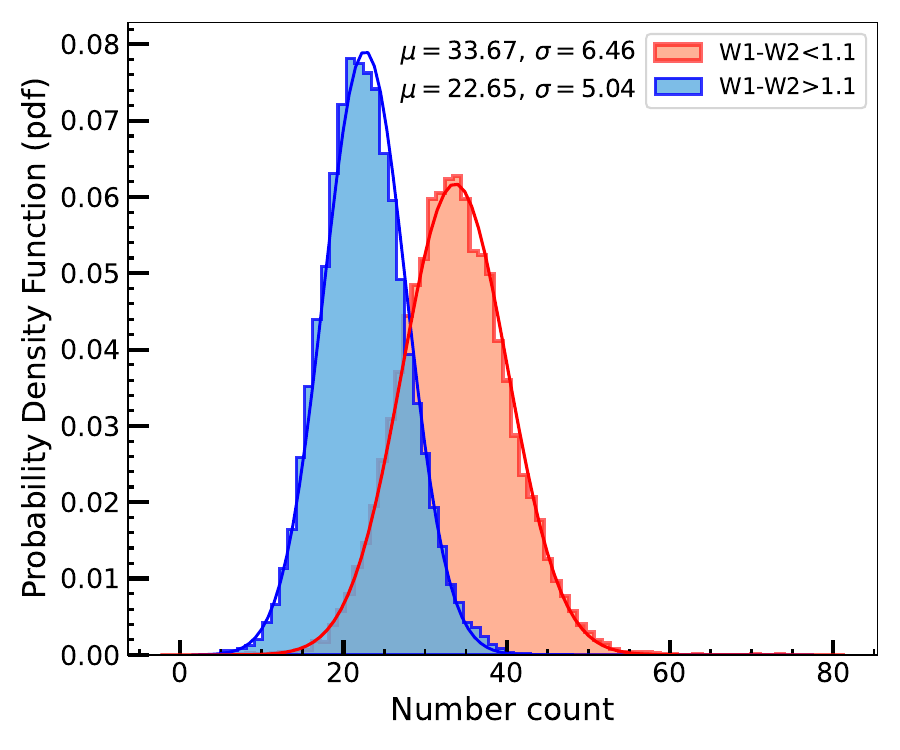}
    \caption{\label{fig:lowhighcolorcutNumberCountDist}The normalised distribution of number counts for low and high color cut data. These plots are well described by Gaussian distributions.}
\end{figure}

The most important observation is that the direction of the dipoles for $W1-W2>1.1$ points approximately opposite to the Galactic center. This indicates that this data set is likely to be very significantly contaminated with the Galactic emissions. We also see this clearly by making a source count map for different cuts. In Figs. \ref{fig:full_data_sky_distribution}, \ref{fig:full_data_sky_distribution_low_color_cut} and \ref{fig:full_data_sky_distribution_high_color_cut} we show the number density i.e., number of sources or number count per deg$^{2}$, map for full data, low color cut data $W1-W2<1.1$ and high color cut data $W1-W2>1.1$ respectively. The number density values on the color bars are chosen in such a way as to show the presence of the dipole and quadrupole signals in the respective data sets. We see that while the low color cut data $W1-W2<1.1$ shows a dipole roughly similar to the full data, the high color cut $W1-W2>1.1$ data shows a completely different behaviour, closely correlated with our galaxy. The spectroscopic redshift distributions for the two cuts are shown in Fig. \ref{fig:lowhighcolorcutredshiftDist.}. We find that the mean redshift for the cut $W1-W2>1.1$ is greater than that for $W1-W2<1.1$. The distributions of number counts for both the cuts are shown in Fig. \ref{fig:lowhighcolorcutNumberCountDist}. We see that both are well described by a Gaussian distribution. 

\begin{figure}
    \includegraphics[width=\columnwidth]{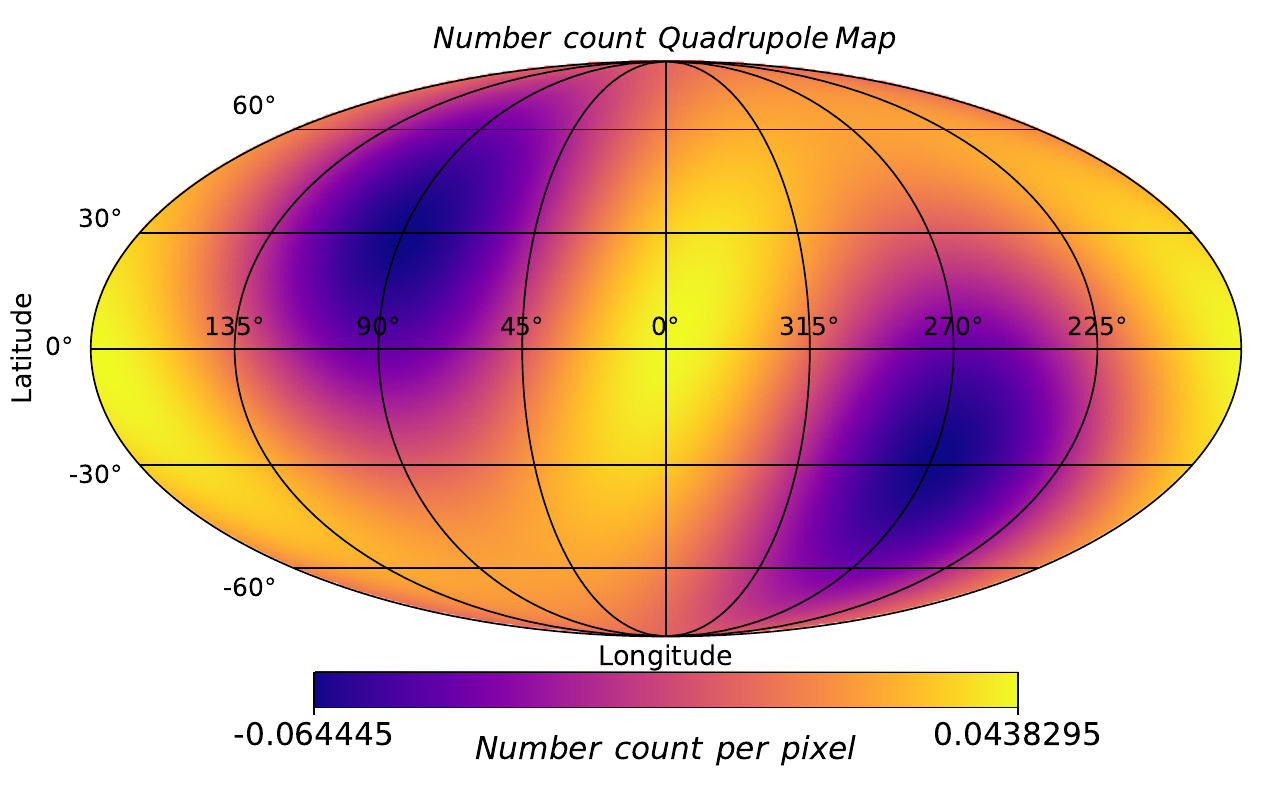}
    \caption{\label{fig:Quadrupole_low_color_cut} Quadrupole map of number count of low color cut $(1.1>W1-W2\ge 0.8)$ data}
\end{figure}

\begin{figure}
    \includegraphics[width=\columnwidth]{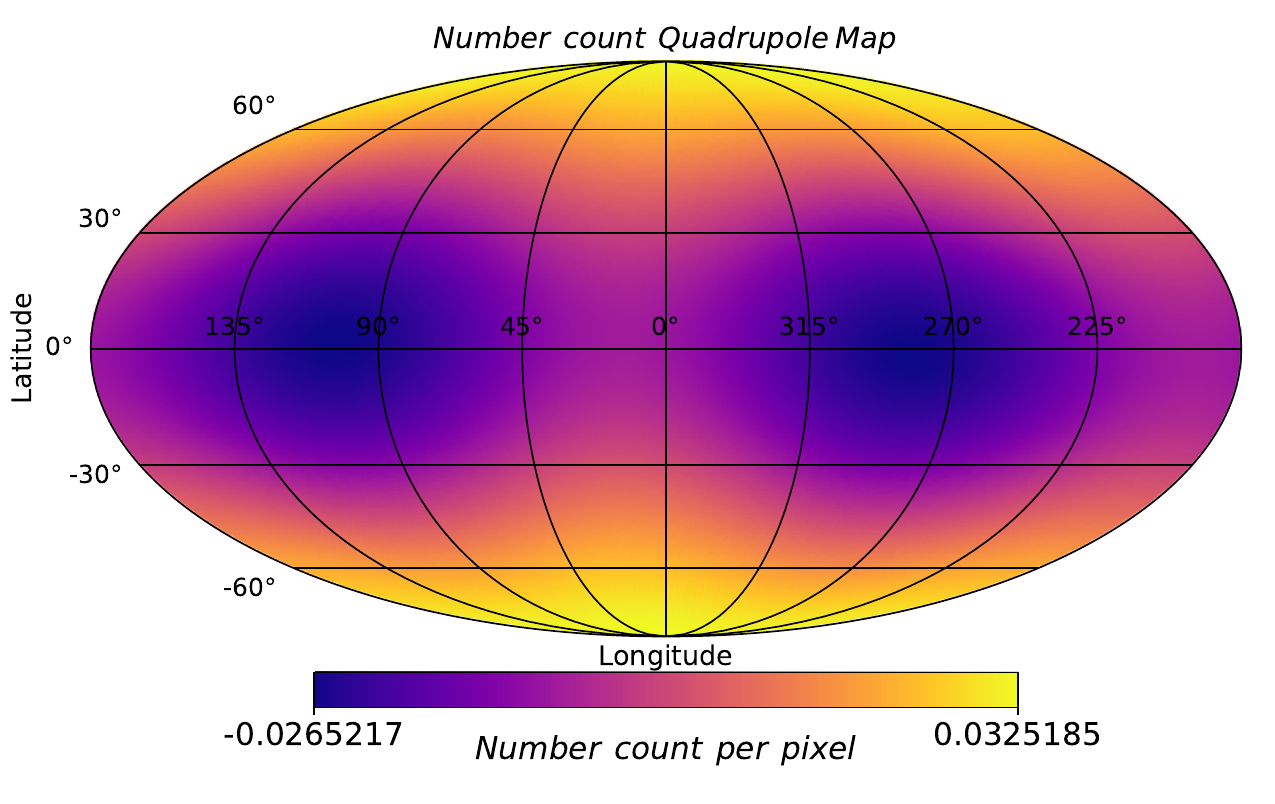}
    \caption{\label{fig:Quadrupole_high_color_cut}  Quadrupole map of number count of high color cut $(1.4>W1-W2>1.1)$ data}
\end{figure}

Figs. \ref{fig:Quadrupole_low_color_cut} and \ref{fig:Quadrupole_high_color_cut} display the fitted quadrupole present in low and high color cut data respectively. In these plots, we show only the quadrupole moment $N_{Q}(\theta,\phi)$ i.e.,
\begin{equation}
    N_{Q}(\theta,\phi) = \frac{N(\theta,\phi)-M_{o}}{M_{o}}\,
\end{equation}
where $N(\theta, \phi)$ is obtained from  Eq. \ref{eq:numbercount_sky_model} by ignoring the dipole parameters $(D_{x}, D_{y}, D_{z})$. In low color cut data, we find that the quadrupole axis of symmetry, i.e., direction of cold spots, points towards the ecliptic poles, which may be correlated with the WISE scanning strategy. The quadrupole moment is nearly the same as in the full data \cite{Kothari_2024}. This fact is also reflected in our complete data map (Fig \ref{fig:full_data_sky_distribution}). In the case of high color data, the quadrupole is closely correlated with the Galactic plane, similar to the dipole signal. In order to investigate the effect of the Galactic plane, we study how the results change as we eliminate the Galactic plane with more stringent cuts starting from $|b|>30^{\circ}$ to $|b|> 50^{\circ}$. The results are shown in Table \ref{table:gal_cut_results}. We see that the results are very similar for this entire range of cuts. For both the color cuts, the magnitude of dipoles as well as their directions show only mild change as we make the Galactic plane cut more stringent. We next explore the possibility that the data corresponding to the cut $W1-W2>1.1$ shows Galactic contamination primarily due to the low flux sources. We study how the dipole signal changes as we increase the low flux cut from $0.085$ mJy to $0.15$ mJy. The results are shown in Table \ref{table:high_color_lowflux_cut_results}. We find that the dipole results are very stable against the value of $S_{cut}$, even after loosing $\sim 50\%$ sources at $S_{cut}=0.15$ mJy. For all the flux cuts used, the quadrupole is also seen to align with the Galactic plane. 

Our results show that there is significant Galactic contamination in the data set corresponding to the color cut $W1-W2>1.1$. The origin of this contamination is unknown and seems to be present even at high galactic latitudes and is not limited to low flux sources. The dust map used to correct for galactic extinction also does not provide any clues to the origin of this effect. For example, scaling the extinction correction in $W2$ band based on this dust map i.e., corrected magnitude $W_{correct} = W_{obs} - fA_{W}$ where $0<f<1$, leads to a systematic change in the extracted quadrupole but has negligible effect on the dipole. Hence, we are unable to identify the source of the dipole in this data set.

Given that the data for $W_1-W_2>1.1$ is unreliable, it is best to use data only $W_1-W_2<1.1$ for extracting the cosmological dipole. We see from Table \ref{table:highlowDipole} that this set also leads to a very significant dipole which is most likely of cosmological origin. However, its direction, particularly the galactic longitude, deviates considerably from that obtained from the full data. Furthermore the extracted velocity turns out to be $900 \pm 113$ Km/s which is significantly larger than what is obtained from the full data. This is primarily due to the smaller value of $\alpha$ for that data set. It is clear that this extracted velocity deviates from the CMB dipole velocity at approximately 4.7 sigma significance. Furthermore, we point out that this set peaks at redshifts of order unity, as seen from Fig. \ref{fig:lowhighcolorcutNumberCountDist}. Hence, the extracted dipole is reliable only for this redshift range.


\begin{table*}
    \centering
    \caption{Dipole results for full data set.}
    \label{table:fulldataDipoleResults}
    \begin{tabular}{p{1.5cm}p{1.5cm}p{2.0cm}p{3.0cm}p{2.0cm}p{1.0cm}}
        \hline
        \noalign{\vskip 0.2cm}
        \# Sources & $x$ & Mean Alpha($\bar{\alpha}$)& Direction $(l^{\circ},\:b^{\circ})$ & $D$& $\chi^2$/dof\\
        \noalign{\vskip 0.2cm}
        \hline
        \noalign{\vskip 0.1cm}
        $1307516$ & $1.579\pm0.001$ & $1.26$ & $225\pm 9,\:\: 31\pm 7$  & $ 0.015\pm 0.002$ & $1.536$\\
        \noalign{\vskip 0.1cm}
        \hline
    \end{tabular}
\end{table*}

\begin{table*}
    \centering
    \caption{Dipole results after employing color cuts.}
    \label{table:2}
    \begin{tabular}{ p{4.0cm} p{2.0cm} p{4.0cm} p{2.5cm} p{1.5cm} } 
        \hline
        \noalign{\vskip 0.2cm}
        Color cut[mag.] & \# Sources & Direction$(l^{\circ},\:\: b^{\circ})$ & $D$ & $\chi^2$/dof \\ 
        \noalign{\vskip 0.2cm}
        \hline
        \noalign{\vskip 0.1cm}
        $0.8 \leq W1-W2<0.9$ & $319812$ & $278\pm 13,\:\: 20\pm 9$  & $ 0.019\pm 0.004$ & $1.144$\\
        \noalign{\vskip 0.1cm}
        $0.9<W1-W2<1.0$ & $244725$ & $293\pm 16,\:\: 23\pm 11$ & $ 0.018\pm 0.005$ & $ 0.959$\\
        \noalign{\vskip 0.1cm}
        $1.0<W1-W2<1.1$ & $210816$ & $315\pm 29,\:\: 32\pm 22$ & $ 0.012\pm 0.005$ & $ 0.846$\\
        \noalign{\vskip 0.1cm}
        $1.1<W1-W2<1.2$ & $180454$ & $209\pm 27,\:\: 43\pm 22$ & $ 0.016\pm 0.005$ & $ 0.708$\\
        \noalign{\vskip 0.1cm}
        $1.2<W1-W2<1.3$ & $134284$ & $187\pm 32,\:\:  9\pm 19$ & $ 0.012\pm 0.006$ & $ 0.505$\\
        \noalign{\vskip 0.1cm}
        $1.3<W1-W2<1.4$ & $62593$  & $205\pm 55,\:\: -12\pm 33$ & $0.010\pm 0.009$ & $ 0.258$\\
        \hline
    \end{tabular}
\end{table*}

\begin{table*}
    \centering
    \caption{Dipole results for color cuts $1.1 > W1-W2 \ge 0.8$ and $1.4>W1-W2>1.1$.}
    \label{table:highlowDipole}
    \begin{tabular}{p{2.8cm} p{2.0cm} p{2.0cm} p{2.0cm} p{2.5cm} p{1.5cm} p{1.2cm}}
        \noalign{\vskip 0.2cm}
        \hline
        Color cut[mag.] & $x$ &Mean Alpha($\bar{\alpha})$ & \# Sources & Direction$(l^{\circ},\:b^{\circ})$ & $D$ & $\chi^2$/dof\\
        \noalign{\vskip 0.2cm}
        \hline
        \noalign{\vskip 0.1cm}
        $1.1 > W1-W2 \ge 0.8$& $1.660 \pm 0.002$ & $0.86$ & $781760$ & $283\pm 11,\:\: 29\pm 8$  & $ 0.016\pm 0.002$ & $1.338$\\
        \noalign{\vskip 0.1cm}
        $1.4>W1-W2>1.1$& $1.473\pm 0.002$ &  $1.86$ & $525708$ & $194\pm 7,\:\: 19\pm 4$ & $ 0.030\pm 0.003$ & $ 1.208$\\
        \hline
        \noalign{\vskip 0.1cm}
    \end{tabular}
\end{table*}

\begin{table*}
    \centering
    \caption{Dipole results of number counts for low and high color cuts for different Galactic plane cuts, $b_{\text{cut}}^\circ$, are presented, and sources with Galactic latitudes $b^{\circ}>|b_{\text{cut}}^\circ|$ are used to extract the dipole signal.}
    \label{table:gal_cut_results}
    \begin{tabular}{p{3.0cm} p{2.0cm} p{2.0cm} p{4.0cm} p{2.0cm} p{1.5cm}} 
        \hline
        \noalign{\vskip 0.2cm}
        Color cut[mag.] & $b_{cut}^{\circ}$ & \# Sources & Direction$(l^{\circ},\:b^{\circ})$& $D$ & $\chi^2$/dof \\
        \noalign{\vskip 0.2cm}
        \hline
        \noalign{\vskip 0.1cm}
        $W1-W2<1.1$& $30$ & $781760$ & $283\pm 11,\:\: 29\pm 8$  & $ 0.016\pm 0.002$ & $1.338$\\
        \noalign{\vskip 0.1cm}
        \empty & $35$ & $683716$ & $285\pm 10,\:\: 23\pm 6$ & $ 0.018\pm 0.003$ & $ 1.336$\\
        \noalign{\vskip 0.1cm}
        \empty & $40$ & $585372$ & $284\pm 16,\:\: 31\pm 11$ & $ 0.014\pm 0.003$ & $ 1.335$\\
        \noalign{\vskip 0.1cm}
        \empty & $45$ & $472725$ & $262\pm 17,\:\: 28\pm 10$ & $ 0.015\pm 0.004$ & $ 1.345$\\
        \noalign{\vskip 0.1cm}
        \empty & $50$ & $383611$ & $248\pm 21,\:\: 26\pm 11$ & $ 0.015\pm 0.005$ & $ 1.343$\\
        \noalign{\vskip 0.1cm}
        $W1-W2>1.1$& $30$ & $525708$ & $194\pm 7,\:\: 19\pm 4$  & $ 0.030\pm 0.003$ & $1.208$\\
        \noalign{\vskip 0.1cm}
        \empty & $35$ & $460534$ & $193\pm 8,\:\: 17\pm 5$  & $ 0.028\pm 0.003$ & $1.202$\\
        \noalign{\vskip 0.1cm}
        \empty & $40$ & $394977$ & $194\pm 11,\:\: 20\pm 6$  & $ 0.024\pm 0.004$ & $1.214$\\
        \noalign{\vskip 0.1cm}
        \empty & $45$ & $319842$ & $191\pm 12,\:\: 19\pm 6$  & $ 0.026\pm 0.005$ & $1.211$\\
        \noalign{\vskip 0.1cm}
        \empty & $50$ & $259920$ & $192\pm 15,\:\: 19\pm 7$  & $ 0.026\pm 0.006$ & $1.182$\\
        \noalign{\vskip 0.1cm}
        \hline
    \end{tabular}
\end{table*}

\begin{table*}
	\centering
	\caption{Dipole results of number count for high color cut$(W1-W2>1.1)$ data with Galactic plane cut ($b>|b_{cut}|=30^{\circ}$) for given lower flux cuts $S_{cut}$ i.e., sources with flux density $>S_{cut}$[mJy] are used to extract dipole signal. The remaining number of sources in percentage is also given for comparison after the lower flux cut. In each case quadrupole lies in the Galactic plane.}
	\label{table:high_color_lowflux_cut_results}
	\begin{tabular}{p{3.8cm} p{2.8cm} p{4.0cm} p{2.5cm} p{1.5cm}} 
		\hline
		\noalign{\vskip 0.2cm}
		$S_{cut}$[mJy] & \# Sources & Direction$(l^{\circ},\:b^{\circ})$& $D$ & $\chi^2$/dof \\ 
		\noalign{\vskip 0.2cm}
		\hline
		\noalign{\vskip 0.1cm}
		$0.085$ & $525708(100\%)$ & $194\pm 7,\:\: 19\pm 4$  & $ 0.030\pm 0.003$ & $1.208$\\
		\noalign{\vskip 0.1cm}
		$0.087$ & $512363(97.46\%)$ & $195\pm 7,\:\: 19\pm 4$ & $ 0.029\pm 0.003$ & $ 1.205$\\
		\noalign{\vskip 0.1cm}
		$0.089$ & $497640(94.66\%)$ & $195\pm 6,\:\: 17\pm 4$ & $ 0.031\pm 0.003$ & $ 1.199$\\ 
		\noalign{\vskip 0.1cm}
		$0.091$ & $483564(91.98\%)$ & $194\pm 7,\:\: 17\pm 4$ & $0.030\pm 0.003$ & $1.200$\\
		\noalign{\vskip 0.1cm}
		$0.093$ & $470116(89.43\%)$ & $192\pm 7,\:\: 18\pm 4$ & $0.031\pm 0.003$ & $1.195$\\ 
		\noalign{\vskip 0.1cm}
		$0.095$ & $457156(86.96\%)$ & $193\pm 7,\:\: 16\pm 4$ & $0.032\pm 0.003$ & $1.191$\\
		\noalign{\vskip 0.1cm}
		$0.098$ & $441644(84.01\%)$ & $194\pm 7,\:\: 17\pm 4$ & $0.031\pm 0.003$ & $1.182$\\
		\noalign{\vskip 0.1cm}
		$0.10$   & $427115(81.25\%)$ & $194\pm 7,\:\: 18\pm 5$ & $0.030\pm 0.003$ & $1.176$\\
		\noalign{\vskip 0.1cm}
		$0.12$  & $332670(63.28\%)$ & $197\pm 9,\:\: 20\pm 6$ & $0.027\pm 0.004$ & $1.128$\\
		\noalign{\vskip 0.1cm}
		$0.15$  & $239085(45.48\%)$ & $203\pm 14,\:29\pm 10$ & $0.023\pm 0.004$ & $0.947$\\
		\hline
		\noalign{\vskip 0.1cm}
	\end{tabular}
\end{table*}

\section{Conclusions}\label{sec:conclusion}
The main result of our paper is that the data in the color bin $1.4>W1-W2>1.1$ suffers from strong contamination due to the Milky Way and hence is not reliable for extracting the cosmological dipole. Both the dipole and quadrupole extracted in this bin are strongly correlated with the galactic plane. The source of this contamination is unknown and we are not able to explain it in terms of known Galactic sources. This is particularly true for the dipole which remains unaffected even if we change the magnitude of the extinction correction based on known sources of galactic extinction. In contrast, the remaining data, corresponding to the cut $1.1>W1-W2 \ge 0.8$, which peaks at redshift approximately equal to $1$, shows a dipole roughly aligned with the CMB dipole direction and hence may be of cosmological origin. This data also shows a quadrupole aligned with the ecliptic poles, as seen in the full data, and hence may be attributed to ecliptic bias. The dipole amplitude in this set is strong and shows a $4.7$ sigma deviation from our local velocity derived from the CMB dipole. Hence the data continues to show a strong deviation from the $\Lambda CDM$ model. We emphasize that it is important to determine the origin of the contamination in the color bin $1.4>W1-W2>1.1$, which corresponds to redshifts greater than $1$, in order to obtain a reliable signal in the full data set.

\section*{Data Availability}
Data used in this article are already available in public domain \cite{Secrest:2020CPQ}. 




\bibliographystyle{mnras}
\bibliography{mnras} 

\bsp	
\label{lastpage}
\end{document}